\documentclass[pre,aps,twocolumn]{revtex4-1}
\usepackage{amsfonts,amssymb,amsmath,bm,color}
\usepackage{graphicx}
\begin{document}

\title{Profile of a coherent vortex in two-dimensional turbulence at static pumping}

\author{I.V.Kolokolov$^{1,2}$ and V.V.Lebedev$^{1,2}$}

\affiliation{$^1$ Landau Institute for Theoretical Physics, RAS, \\
142432, Chernogolovka, Moscow region, Russia; \\
$^2$ National Research  University Higher School of Economics, \\
101000, Myasnitskaya ul. 20, Moscow, Russia.}

\begin{abstract}

We examine the velocity profile of coherent vortices appearing as a consequence of the inverse cascade of two-dimensional turbulence in a finite box in the case of static pumping. We demonstrate that in the passive regime the flat velocity profile is realized, as in the case of pumping short correlated in time. However, in the static case the energy flux to large scales is dependent on the system parameters. We demonstrate that it is proportional to $f^{4/3}$ where $f$ is the characteristic force exciting turbulence.

\end{abstract}

\maketitle

Turbulence is a chaotic state of fluid motions realized at large Reynolds numbers $\mathrm{Re}$, see, e.g., \cite{Monin}. In a number of cases turbulence appears to be effectively two-dimensional 
\cite{BE12}. Already first theoretical works \cite{67Kra,68Lei,69Bat} devoted to two-dimensional turbulence reveal its principal difference from three-dimensional one. The difference is related to existence of two quadratic quantities (energy and enstrophy) conserved by two-dimensional Euler equation. That leads to two different cascades produced by the non-linear interaction: enstrophy is carried from the pumping scale to smaller scales (direct cascade) whereas energy is carried to larger scales (inverse cascade). The enstrophy is dissipated due to viscosity at scales smaller than the pumping length and the energy is dissipated due to bottom friction at scales larger than the pumping length.

Statistical properties of the velocity fluctuations in the inverse cascade were investigated both experimentally \cite{Tabeling} and numerically \cite{BCV}. Results of the works are in a good agreement with the analytical theory developed for an unbound system \cite{KraM}. Particularly, the normal Kolmogorov scaling is observed in the inverse cascade \cite{BE12}. The normal scaling in the inverse cascade of two-dimensional turbulence is in contrast with the anomalous scaling observed in three-dimensional turbulence \cite{Frisch}. Some theoretical arguments in favor of the normal scaling in the inverse cascade were presented in the work \cite{Fal09}, where the feature was related to a leading role of the converging Lagrangian trajectories in the inverse cascade.

In an unbound two-dimensional system the inverse cascade is terminated at a scale $L_\alpha$, which is determined by a balance between the energy flux $\epsilon$ (per unit mass) toward large scales and the bottom friction. Using the Kolmogorov estimate $(\epsilon L_\alpha)^{1/3}$ for the velocity at the scale $L_\alpha$, one obtains from the balance $L_\alpha=\epsilon^{1/2}\alpha^{-3/2}$, where $\alpha$ is the bottom friction coefficient. If the box size $L$ is smaller than $L_\alpha$, then the energy carried by the inverse cascade to scales of the order of the box size $L$ starts to accumulate there. The scenario is realized if the pumping scale $l$ is much less than the box size $L$, $l\ll L$, otherwise there is no space for the inverse cascade carrying the energy to the scale $L$.

The accumulation of the energy at the scale $L$ leads to appearing an intensive large-scale motion including big coherent vortices. The tendency towards the formation of the vortices was indicated already in the first works devoted to two-dimensional turbulence, both experimental \cite{Somm} and numerical \cite{Smith1,Smith2, Borue}. It was established in the numerical work \cite{Colm}, where periodic boundary conditions are utilized, that due to the inverse cascade in a square box a stable structure appears that is the vortex dipole. A bit different coherent vortex structure is generated in laboratory experiments in a square box \cite{Shats,Xia2}. From the theoretical point of view, the last case corresponds to zero boundary conditions at the box boundaries for the flow velocity.  Formation of  vortices in a rectangular geometry was observed numerically in Ref. \cite{FLF17}.

The work \cite{14LBFKL} reports results of numerical simulations of two-dimensional turbulence for a periodic setup in a square box with pumping short correlated in time. The average velocity profile of the coherent vortex appears to be highly isotropic in a range of separations from the vortex center. In the same work an interval of separations from the vortex center was found where the flat velocity profile is realized and some theoretical arguments toward the flat profile were formulated. In our paper \cite{KL16-1} we demonstrated that the flat profile corresponds to the passive (quasilinear) regime of turbulent fluctuations, and in  \cite{KL16-2} we calculated the structure function of the velocity fluctuations. In all the works the model of the pumping short correlated in time was used. The question is about universality of the results obtained in the framework of the model and their applicability to experimental observations.

Here we analyze the case where two-dimensional turbulence is excited by a static (time-independent) pumping. It is characterized by a ``frozen'' force $\bm f(x,y)$ (per unit mass) applied to a thin fluid film. The pumping correlation length $l$ is assumed to be much smaller than the box size $L$. Particularly, one can think about the force $f(x,y)$, periodic in space with the period $l$. The static case corresponds to the setup of the experimental works \cite{Tabeling,Shats} where Lorentz forces are used to excite turbulence. The inverse cascade is formed if the pumping force is strong enough. Namely, the inequality
 \begin{equation}
 f \gg l \alpha^2,
 \label{basicc}
 \end{equation}
should be satisfied. Below we assume that the inequality (\ref{basicc}) is valid.

We consider the case of a finite box with the size $L$ satisfying the inequality $L\ll L_\alpha$. In the case the inverse cascade carries the energy upto scales of the order of the box size $L$ where the energy is accumulated until a balance between the energy flux and its dissipation is reached. That leads to formation of a large-scale flow with a characteristic velocity $V$ that can be evaluated as follows. The energy dissipation in the system is estimated as $\alpha V^2$. In the stationary case the dissipation should be equal to the energy flux, $\epsilon\sim \alpha V^2$. Thus $V\sim \epsilon^{1/2}\alpha^{-1/2}$. Note that $V\gg (\epsilon L)^{1/3}$ due to the inequality $L\ll L_\alpha$. By other words, the large-scale motion in the finite box is much more intensive than motions of the scale $L$ in the inverse cascade in an unbound system.

The large-scale flow in a finite box consists of a number of strongly interacting modes. In this case one expects a chaotic behavior \cite{InRo}. To examine small-scale motions in the case, one passes to the reference frame attached to the large-scale motion. In the reference frame the pumping force $\bm f$ is time-dependent and has an effective correlation time $\tau=l/V$. Then the energy production rate can be estimated as $\epsilon\sim f^2 \tau$. Equating the value to the energy dissipation $\alpha V^2$, we arrive at the estimates (see Ref. \cite{KL17})
 \begin{eqnarray}
 V\sim f^{2/3} l^{1/3} \alpha^{-1/3},
 \label{lveloc} \\
 \epsilon\sim f^{4/3} l^{2/3} \alpha^{1/3},
 \label{epsilon}
 \end{eqnarray}
for the large-scale velocity $V$ and the energy flux $\epsilon$.

Let us stress that for the static pumping the energy flux (\ref{epsilon}) is dependent on $\alpha$. The dependence is a bit counter-intuitive: at decreasing $\alpha$ the energy flux diminishes. This is a consequence of the suppression of the energy production by the large-scale flow the velocity of which (\ref{lveloc}) grows as $\alpha$ diminishes. Indeed, the effective correlation time $\tau=l/V$ of the pumping diminishes as the large-scale velocity $V$ grows and therefore the energy production $\sim f^2 \tau$ diminishes. Similar conclusions are presented in the recent paper \cite{Anna17}.

In contrast to the pumping short correlated in time, where the energy production rate is prescribed, in our case of the ``frozen'' force the energy production is determined self-consistently. The factor in the estimate  (\ref{epsilon}) depends on details of the large-scale flow statistics. Moreover, the energy production rate can be inhomogeneous. Say, the rate can differ inside the coherent vortices and at their periphery. However, as we demonstrate below, the estimate  (\ref{epsilon}) works everywhere, irrespective to the region of the box.

The estimate $\epsilon\sim f^2 \tau$ implies that the pumping effective correlation time $\tau$ is the smallest characteristic time at the pumping scale. One has to check the condition. There are two different cases. In the first case the flow dynamics at the pumping scale is non-linear. Then the fluctuation velocity at the pumping scale is estimated as $(\epsilon l)^{1/3}$ and therefore the nonlinear rate is estimated as $\Gamma= \epsilon^{1/3} l^{-2/3}\sim \alpha^{1/9}l^{-4/9}f^{4/9}$. Using the estimates (\ref{lveloc},\ref{epsilon}) one can easily check that the inequality $\tau\Gamma\ll 1$ is equivalent to the basic condition (\ref{basicc}). The second passive (quasilinear) case can be realized inside the coherent vortices, it is analyzed below.

In addition to the smooth large-scale motion with the characteristic velocity (\ref{lveloc}), in the finite box there appear coherent structures that are big vortices \cite{Colm,Shats,14LBFKL}. Therefore a question appears about an internal structure of the vortices. For a short-correlated in time pumping, the problem was investigated in the works \cite{14LBFKL,KL16-1,KL16-2}. Let us analyze the structure of the coherent vortex in the static case. We assume that (in the reference frame attached to the vortex center) the velocity profile of the vortex is isotropic and can be, consequently, characterized by a polar velocity $U$ dependent on the separation $r$ from an observation point to the vortex center.

The motions of the vortex center are determined by the large-scale flow and, consequently, the characteristic velocity of the vortex center can be estimated as (\ref{lveloc}). Since the large-scale motion is chaotic, the pumping in the reference frame attached to the vortex center is chaotic as well, it can be characterized by the amplitude $f$ and the correlation time $\tau=l/V$. Thus, one should generalize the procedure of the works  \cite{KL16-1,KL16-2} to the case of a random pumping with a finite correlation time.

To find the velocity profile of the coherent vortex in the reference system attached to the vortex center one produces the Reynolds decomposition. In the reference frame the large-scale flow is isotropic and is characterized by the polar velocity $U(r)$, where $r$ is separation from the vortex center. Let us designate the radial and the polar components of the fluctuating velocity as $v$ and $u$, respectively. Then the equation on $U$ is
 \begin{equation}
 \alpha U+(\partial_r +2/r)\langle u v \rangle=0,
 \label{averu}
 \end{equation}
where angular brackets designate time averaging. The equation (\ref{averu}) is a direct consequence of Euler equation with addition of the bottom friction term \cite{14LBFKL,KL16-1,KL16-2}.

To find the average $\langle u v \rangle$ in Eq. (\ref{averu}), one should examine small-scale fluctuations with scales of order $l$ where the average is formed. Following the works \cite{KL16-1,KL16-2} we consider the region inside the vortex where the small-scale flow fluctuations are passive, the approximation is usually called quasilinear. It is convenient to start from the equation for the small-scale vorticity $\varpi$ that in the passive case can be written as
 \begin{equation}
 \partial_t\varpi+ (U/r)\partial_\varphi \varpi
 +\hat\Gamma\varpi= \phi,
 \label{vorteq}
 \end{equation}
where $\phi=\mathrm{curl}\, \bm f$,  $\varphi$ is the polar angle and $\hat\Gamma\varpi$ is the dissipation term.

We examine the flow fluctuations near a radius $r_0\gg l$. Then one can expand in Eq. (\ref{vorteq}) the velocity $U$ in the series in $r-r_0$ and we find
 \begin{equation}
 \partial_t \varpi+ (U/r_0)\partial_\varphi \varpi
 +\Sigma (r-r_0) \partial_\varphi \varpi
 +\hat\Gamma \varpi=\phi,
 \label{varpieq}
 \end{equation}
where both, $U$ and $\Sigma$, are taken at $r=r_0$. The quantity
 \begin{equation}
 \Sigma= r\partial_r(U/r)
 =\partial_r U -U/r.
 \label{shear}
 \end{equation}
is the effective shear rate of the average velocity.

To find $U$ from the equation (\ref{averu}) one should solve the equation (\ref{varpieq}) and then find the average $\langle u v \rangle$. For the pumping short correlated in time, the program was realized in Ref. \cite{KL16-1}. In contrast to the case, for the pumping with finite correlation time the term with the average velocity $U$ in Eq. (\ref{varpieq}) doesn't drop from the answer for the average $\langle u v \rangle$. The corresponding analysis is done in Appendix. Below, we use results of the analysis.

As it is demonstrated in Appendix, if $U\sim V$, then $\langle uv \rangle \sim   \epsilon/\Sigma$, see Eq. (\ref{epsilon2}). Substituting the result into Eq. (\ref{averu}) one obtains
 \begin{equation}
 U\sim (\epsilon/\alpha)^{1/2}\sim
 f^{2/3} l^{1/3} \alpha^{-1/3},
 \label{estimu}
 \end{equation}
that is $U\sim V$, indeed. That justifies our approach. Note that the estimation for $U$ (\ref{estimu}) is the same as in the case of pumping short correlated in time. However, in the static case one cannot explicitly establish the factor in the relation $U\sim (\epsilon/\alpha)^{1/2}$.

Let us find the upper boundary $R_u$ of the passive region. Equating the shear rate $\Sigma$ to the nonlinear rate at the pumping scale $\epsilon^{1/3} l^{-2/3}$, one finds for the boundary radius
 \begin{equation}
 R_u \sim \epsilon^{1/6} \alpha^{-1/2} l^{2/3}
 \sim f^{2/9} l^{7/9} \alpha^{-4/9}.
 \label{vorbound}
 \end{equation}
The quantity is independent of $L$, as in the case of pumping short-correlated in time \cite{KL16-1}. Note that $R_u\gg l$ as a consequence of the condition (\ref{basicc}). The quantity $R_u$ can be interpreted as the radius of the big coherent vortex.

In the passive region $r<R_u$ the dissipation rate $\Gamma$ at the pumping scale can be found as follows. The eddy diffusivity there is estimated as $\epsilon/\Sigma^2$. Therefore the characteristic rate is $\Gamma\sim \epsilon \Sigma^{-2} l^{-2}$. One can easily check, using the estimates (\ref{lveloc},\ref{epsilon}) and the inequality (\ref{basicc}), that $\Gamma<\Sigma$ at $r<R_u$ and $\Sigma$ becomes of order $\Sigma$ at $r=R_u$. Note that the estimate $\Gamma\sim \epsilon \Sigma^{-2} l^{-2}$ is equivalent to $\Gamma\sim \alpha (r/l)^2$.

Now we can finish proof of self-consistency of our approach. Since $\Gamma<\Sigma$ in the passive region, one can check that the condition $\Sigma\tau\ll 1$ is equivalent to the inequality $r\gg l$. Next, $\Gamma\sim \alpha (r/l)^2$ if $r\gg l$. Therefore the bottom friction $\alpha$ can be neglected at the pumping scale $l$. In the opposite case one would encounter the situation noted in the work 
\cite{KL16-1}: if the dissipation at the pumping scale is dispersionless (that is independent of the wave vector $q$) then $\langle uv \rangle=0$. Fortunately, we avoid the scenario.

To conclude, we established properties of coherent vortices appearing in stationary two-dimensional turbulence excited in a finite box by a static pumping. There is an interval of separations from the vortex center where the coherent polar velocity of the vortex has the flat profile, that is independent of the separation. It is interesting that the polar velocity has the same value as the characteristic velocity of the large-scale flow outside the vortices. The interval of separations where the flat profile has to be observed is restricted by the radius (\ref{vorbound}) and corresponds to the passive (quasilinear) regime of the small-scale flow fluctuations.

Let us stress that the unusual dependencies of the characteristic velocity $V$ (\ref{lveloc}) and of the energy flux (\ref{epsilon}) on the pumping force $f$ are related to the static (``frozen'') character of the pumping. Thus the case differs from the one considered in the works \cite{KL16-1,KL16-2}, where the pumping short correlated in time, was analyzed. However, the final conclusions concerning the existence of the scale interval with the flat coherent velocity profile with the value of the order of the velocity of the large-scale motion is the same. Thus the features can be treated as universal aspects of two-dimensional turbulence independent of the pumping details.

We thank G. Falkovich, A. Levchenko and A. Orlov for valuable discussions. The work is supported by RScF grant 14-22-00259.

\appendix
\section{}\label{appA}

Producing Fourier-transform in terms of the variables $x_1=r-r_0$, $x_2=r_0\varphi$, one obtains from Eq. (\ref{varpieq})
 \begin{equation}
 \partial_t\varpi({\bm{k}})+\mathrm{i}k_2 U \varpi({\bm{k}})
 -\Sigma k_2\frac{\partial\varpi({\bm{k}})}{\partial k_1}
 +\Gamma\varpi({\bm{k}})=\phi({\bm k}).
 \label{pass1}
 \end{equation}
A solution of the equation (\ref{pass1}) is
 \begin{eqnarray}
 \varpi_{\bm{k}}(t)
 =\int^t d\tau\, e^{-\mathrm{i}k_2U(t-\tau)}
 \phi[k_1+(t-\tau)\Sigma k_2,k_2]\qquad
 \label{passolv} \\
 \times\exp\left\{-\int\limits_\tau^t\,d\tau'\Gamma
 \left(\sqrt{\left[k_1+(t-\tau')\Sigma k_2\right]^2+k_2^2}\right)\right\},
 \nonumber
 \end{eqnarray}
where integration is performed over the time of active pumping. For the stationary case, the integration interval over $\tau$ is $(-\infty,t)$.

Thus the vorticity pair correlation function is expressed in terms of the pair correlation function of the pumping $\bm f$. At small scales (of order $l$) its statistical properties are homogeneous, that is the pair correlation function of $\bm f$ depends on the differences of the coordinates. It is also isotropic. Thus
 \begin{equation}
 \langle \phi(t,{\bm{k}}) \phi(t',{\bm{q}})\rangle
 = (2\pi)^2\delta(\bm{k}+\bm{q})k^2\Phi(t-t, k),
 \label{pumpfi}
 \end{equation}
where the characteristic value of $k$ is $l^{-1}$ and
 \begin{equation}
 \int dt\, \frac{d^2\bm{k}}{(2\pi)^2}\Phi(t,k)\sim \epsilon.
 \label{epsilon1}
 \end{equation}

Below, the relaxation factor that is the last exponent in Eq. (\ref{passolv}) will be substituted by unity. The reason is that the inequality $\Sigma\gg \Gamma$ is valid and all integrals discussed further converge without the factor. However, to conduct calculations, one should change order of integrations. The procedure is justified for the infinite integration interval just by presence of the dissipation factor.

In the chosen reference system components of the velocity are $(v,u)$. Using the relations between the Fourier components
\begin{equation*}
u({\bm k})=-\mathrm{i}\frac{k_1}{{\bm k}^2}\varpi({\bm{k}}),\quad
v({\bm k})=\mathrm{i}\frac{k_2}{{\bm k}^2}\varpi({\bm{k}}),
\end{equation*}
one obtains
 \begin{eqnarray}
 \langle uv \rangle = \int \frac{d^2 {k} d^2 {q}}{(2\pi)^4}\langle u({\bm k})v({\bm q})\rangle=
 \nonumber \\
 -\int_0^\infty d\tau_1\int_0^\infty d\tau_2
 \int \frac{d^2 q}{(2\pi)^2}\,q^2
 \Phi(\tau_1-\tau_2,q)
 \nonumber \\
 \times\frac{e^{-\mathrm{i}q_2U(\tau_1-\tau_2)}\left(q_1-q_2\Sigma(\tau_1+\tau_2)/2\right)q_2}
 {\left[(q_1-\Sigma\tau_1 q_2)^2+q_2^2\right]
 \left[(q_1-\Sigma\tau_2 q_2)^2+q_2^2\right]}.
\label{uvint}
 \end{eqnarray}
The expression (\ref{uvint}) can be transformed as
 \begin{eqnarray}
 \langle uv \rangle =
-\int_0^\infty d\tau_1\int_0^\infty d\tau_2 \frac{\Phi(\tau_1-\tau_2,q)}{2\Sigma(\tau_1-\tau_2)}
 \nonumber \\
 \int \frac{d^2 q}{(2\pi)^2}\,q^2
 e^{-\mathrm{i}q_2U(\tau_1-\tau_2)}
 \nonumber \\
\left\{\frac{1}{(q_1-\Sigma\tau_1 q_2)^2+q_2^2}-\frac{1}{(q_1-\Sigma\tau_2 q_2)^2+q_2^2}\right\}.
\label{uvint1}
 \end{eqnarray}
Using a symmetry of the integrand, the integral over $\tau_1$ and $\tau_2$ can be rewritten as
\begin{equation*}
\int_0^\infty d\tau_1\int_0^\infty d\tau_2 \to2\int_0^\infty d\tau_1\int_{\tau_1}^\infty d\tau_2.
\end{equation*}

Let us now pass to the variables $\eta=\tau_2-\tau_1$, $\lambda=\tau_1/\eta$. Then one obtains
 \begin{eqnarray}
 \langle uv \rangle
=\frac{1}{\Sigma}
\int \frac{d^2 q}{(2\pi)^2}\,q^2
 \nonumber \\
\int_0^\infty d\eta\,
 e^{-\mathrm{i}q_2U\eta}\Phi(\eta,q)
\int_0^1 \frac{d\lambda}{(q_1-\Sigma\lambda\eta q_2)^2+q_2^2}.
\label{uvdvls1}
 \end{eqnarray}
Since the integrand is invariant under the transformation $q_2,\eta \rightarrow -q_2,-\eta$, one can extend the integration over $\eta$ from $-\infty$ to $+\infty$, and after that one restrict the integration over $q_2$ by positive $q_2$. Next, we pass to the Fourier transform $F(\omega)=\int {dt} \exp(i\omega t) \Phi(t)$, and take the integral over $\eta$ to arrive at
 \begin{eqnarray}
 \langle uv \rangle
=\frac{1}{4\pi \Sigma}
\int\limits_0^\infty \frac{d q_2}{q_2}\int\limits_{-\infty}^{\infty}dq_1\,q^2
\int\limits_{-\infty}^{\infty}\frac{ds}{2\pi}
 \nonumber \\
\int_0^1 d\lambda\,F(s\Sigma \lambda q_2+U q_2,q)
 e^{-q_2|s|+\mathrm{i}q_1 s}.
\label{uvdvls2}
 \end{eqnarray}
where we substituted $\omega=s\Sigma \lambda q_2+U q_2$.

The integral over $s$ is determined by $s\sim l$, therefore $s\Sigma \lambda \sim Ul/r\ll U $. Thus, the first term in the first argument of  $F$ in Eq. (\ref{uvdvls2}) can be neglected. Then the integral over $s$ can be explicitly calculated and we obtain
 \begin{eqnarray}
 \langle uv \rangle
 =\frac{1}{4\pi^2 \Sigma}
 \int\limits_0^\infty d q_2\int\limits_{-\infty}^{\infty}dq_1\,
 F(U q_2,q).
 \label{uvdvls3}
 \end{eqnarray}
Since the pumping correlation time is estimated as $l/V\sim l/U$, the wave vectors $q_1$ and $q_2$ are estimated as $l^{-1}$ and we arrive at the estimation
 \begin{equation}
 \langle uv \rangle\sim \frac{1}{\Sigma}\int d^2q\ F(0,q)
 \sim   \epsilon/\Sigma,
 \label{epsilon2}
 \end{equation}
where we used the relation (\ref{epsilon1}).


\begin{thebibliography}{99}

\bibitem{Monin}
Monin,  A.S \& Yaglom,  A.M. 1971
 Statistical Fluid Mechanics: Mechanics of Turbulence, ( Edited by J.L.Limley)
{\it  Dover Publications, Inc., Mineola, New York, 1971}


  \bibitem[Boffetta \& Ecke (2012)]{BE12}
 {\sc Boffetta, G. \&  Ecke, R.E.} 2012
{Two-Dimensional Turbulence}
 {\it Ann. Rev. Fluid Mech.} {\bf 44}, 427-451.

 \bibitem[ Kraichnan (1967)]{67Kra}
{\sc   Kraichnan, R.~H.} 1967
{Inertial ranges in two-dimensional turbulence}
{\it  Phys. Fluids} {\bf 10}, 1417-1423.


\bibitem[Leith (1968)]{68Lei}
 {\sc  Leith, C.~E.} 1968
{Diffusion approximation for two-dimensional turbulence} {\it  Phys. Fluids} {\bf 11}, 671-673.


 \bibitem[ Batchelor (1969)]{69Bat}
{\sc  Batchelor, G.~K.} 1969
{Computation of the energy spectrum in homogeneous two-dimensional turbulence}
{\it  Phys. Fluids} {\bf 12}, 233-239.

 

 \bibitem[Tabeling (2002)]{Tabeling}
 {\sc Tabeling, P.} 2002
{Two-dimensional turbulence: a physicist approach}
{\it Physics Reports} {\bf 362}, 1.

\bibitem[  Boffetta  et al (2000)]{BCV}
 {\sc   Boffetta, G.,  Celani, A.  \&  Vergassola, M.} 2000
        {Inverse energy cascade in two-dimensional turbulence: Deviations from Gaussian behavior}
        {\it  Phys. Rev. E} {\bf 61}, R29.

 \bibitem[ Kraichnan  \&  Montgomery (1980)]{KraM}
{\sc   Kraichnan, R. H. \&  Montgomery, D.} 1980
{Two-dimensional turbulence}
 {\it Rep. Prog. Phys.} {\bf 43}, 547.

\bibitem[Frisch (1995)]{Frisch}
 {\sc Frisch, U.} 1995
 {Turbulence: The Legacy of A. N. Kolmogorov,
 Cambridge University Press, Cambridge}.


 \bibitem[Falkovich (2009)]{Fal09}
 {\sc Falkovich, G.} 2009  
{Symmetries of the turbulent state}
{\it J. Phys. A: Math. Theor.} {\bf 42}, 123001.

 \bibitem[Sommeria (1986)]{Somm}
 {\sc Sommeria, J.} 1986
{Experimental study of the two-dimensional inverse energy cascade in a square box}
 {\it J. Fluid Mech.} {\bf 170}, 139-168.


\bibitem[ Smith \&  Yakhot(1993)]{Smith1}
{\sc  Smith,  L. M. \&  Yakhot, V.} 1993
{Bose-condensation and small-scale structure generation in a random force driven 2d turbulence}
{\it  Phys. Rev. Lett.} {\bf 71}, 352-355.

\bibitem[ Smith \&  Yakhot(1994)]{Smith2}
{\sc  Smith,  L. M. \&  Yakhot, V.} 1994
{ Finite-size effects in forced, two-dimensional turbulence}
{\it   J. Fluid  Mech.} {\bf  274}, 115-138.

 \bibitem[Borue (1994)]{Borue}
{\sc Borue, V.} 1994
{ Inverse energy cascade in stationary two-dimensional homogeneous turbulence}
{\it  Phys. Rev. Lett.} {\bf 72}, 1475 - 1478.


 \bibitem[Chertkov et al (2007)]{Colm}
{\sc  Chertkov, M.,  Connaughton, C., Kolokolov, I. \& Lebedev, V.} 2007 
{Dynamics of Energy Condensation in Two-Dimensional Turbulence}
{\it  Phys. Rev. Lett} {\bf 99}, 084501.

\bibitem[Xia et al (2009)]{Shats}
{\sc  Xia, H., Shats, M. \&  Falkovich, G.} 2009
{Spectrally condensed turbulence in thin layers}
 {\it Phys. Fluids} {\bf 21}, 125101.

 \bibitem[Francois et al (2014)]{Xia2}
{\sc   Francois, N.,  Xia, H.,  Punzmann, H.,  Ramsden S. \&  Shats M.} 2014
{Three-Dimensional Fluid Motion in Faraday Waves: Creation of Vorticity and Generation of Two-Dimensional Turbulence}
{\it  Phys. Rev. X} {\bf 4}, 021021.
 

\bibitem[Frishman et al (2017)]{FLF17}
 {\sc Frishman, A., Laurie, J.\& Falkovich, G., } 2017
{ Jets or vortices - what flows are generated by an inverse turbulent cascade?}
 {\it  Phys. Rev. Fluids} {\bf 2}, 032602.


\bibitem[ J.~Laurie et al (2014)]{14LBFKL}
 {\sc  Laurie, J., Boffetta, G., Falkovich, G., Kolokolov, I., \& Lebedev, V.} 2014
{Universal Profile of the Vortex Condensate in Two-Dimensional Turbulence,}
{\it  Phys. Rev. Lett.} {\bf 113}, 254503.



\bibitem[ Kolokolov \&  Lebedev (2016A)]{KL16-1}
 {\sc  Kolokolov, I.V. \&  Lebedev, V.V.} 2016
        {Structure of coherent vortices generated by the inverse cascade of two-dimensional turbulence in a finite box}
        {\it  Phys. Rev. E} {\bf 93}, 033104.

				
\bibitem[ Kolokolov \&  Lebedev (2016B)]{KL16-2}
 {\sc  Kolokolov, I.V. \&  Lebedev, V.V.} 2016
        {Velocity statistics inside coherent vortices generated by the inverse cascade of $2d$ turbulence}
        {\it J. Fluid Mech.} {\bf 809}, R2-1.


\bibitem[ Infeld \& Rowlands (2000)]{InRo}
{\sc Infeld, E. \& Rowlands, G.} 2000
 {Nonlinear waves, solitons and chaos, Ch. 11}
{\it  Cambridge Univ. Press}
 

\bibitem[ Kolokolov \&  Lebedev (2017)]{KL17}
 {\sc  Kolokolov, I.V. \&  Lebedev, V.V.} 2017
{ Large-scale flow in two-dimensional turbulence at static pumping }
{\it Pis'ma v ZhETF} {\bf 106}   [{\it JETP Letters}  {\bf 106}].
 
\bibitem[ Frishman  (2017)]{Anna17}
 {\sc  Anna Frishman} 2017
{ The culmination of an inverse cascade:mean flow and fluctuations }
{\it arXiv:1710.09250v1 [physics.flu-dyn]}].
 
\end{thebibliography}
\end{document}